\documentclass[prd,twocolumn,showpacs,superscriptaddress,floatfix,longbibliography]{revtex4-1}
\usepackage{graphicx,amsfonts,amssymb,amsmath,xspace,dsfont}
\usepackage[colorlinks=true,citecolor=green,linkcolor=red,urlcolor=blue]{hyperref}
\usepackage{subfigure}

\usepackage{color}
\definecolor{g}{rgb}{.1,0.4,.1} 
\definecolor{b}{rgb}{0,0.2,1}
\definecolor{rouge}{rgb}{0.82,0.,0.}
\definecolor{vert}{rgb}{0.,0.82,0.}
\definecolor{orange}{rgb}{1,0.5,0.}
\definecolor{bleu}{rgb}{0.,0.,0.82}
\definecolor{m}{rgb}{0.82,0.,0.82}
\definecolor{vert2}{rgb}{0.,0.5,0.}
\definecolor{rougeclair}{rgb}{1.0,0.7,0.7}

\newcommand{\beq}{\begin{equation}}
\newcommand{\be}{\begin{equation}}
\newcommand{\beqn}{\begin{eqnarray}}
\newcommand{\eeq}{\end{equation}}
\newcommand{\ee}{\end{equation}}
\newcommand{\eeqn}{\end{eqnarray}}

\newcommand{\bem}{\begin{pmatrix}}
\newcommand{\eem}{\end{pmatrix}}

\newlength{\ldag}
\newcommand{\bn}{{\mathbf{n}}}

\begin{document}

\title{Parity of Chern numbers in the Kitaev honeycomb model and the sixteenfold way}

\author{Jean-No\"el Fuchs}
\email{fuchs@lptmc.jussieu.fr}
\affiliation{Sorbonne Universit\'e, CNRS, Laboratoire de Physique Th\'eorique de la Mati\`ere Condens\'ee, LPTMC, F-75005 Paris, France}

\author{Sourabh Patil}
\email{sourabhpatil.9@gmail.com}
\affiliation{Indian Institute of Science Education and Research, Pune 411008 India}
\affiliation{Sorbonne Universit\'e, CNRS, Laboratoire de Physique Th\'eorique de la Mati\`ere Condens\'ee, LPTMC, F-75005 Paris, France}

\author{Julien Vidal}
\email{vidal@lptmc.jussieu.fr}
\affiliation{Sorbonne Universit\'e, CNRS, Laboratoire de Physique Th\'eorique de la Mati\`ere Condens\'ee, LPTMC, F-75005 Paris, France}

\begin{abstract}

In two dimensions, topological phases of free Majorana fermions coupled to a $\mathbb{Z}_2$ gauge field are known to be classified according to the Chern number $\nu \in \mathbb{Z}$. Its value mod 16 specifies the type of anyonic excitations. In this paper, we investigate triangular vortex configurations (and their dual) in the Kitaev honeycomb model and show that fourteen of these sixteen phases can be obtained by adding a time-reversal symmetry-breaking term. Missing phases are $\nu=\pm 7$. More generally, we prove that any periodic vortex configuration with an odd number of vortices per geometric unit cell can only host even Chern numbers whereas odd Chern numbers can be found in other cases.
\end{abstract}

\maketitle

%
\section{Introduction}
%
%
Classifying gapped free-fermion systems in terms of topological invariants has been one of the major progresses in the understanding of quantum condensed matter~\cite{Schnyder08,Kitaev09,Ryu10,Chiu16}. At fixed space dimension, this classification scheme, known as the tenfold way, is based  on the analysis of time-reversal, particle-hole, and chiral symmetries which completely specify the topological invariant needed to characterize the corresponding quantum phase. For instance, in two dimensions, integer quantum Hall systems are class-A topological insulators characterized by a Chern number $\nu \in \mathbb{Z}$ which gives the quantized Hall electric conductance~\cite{Thouless82,Kohmoto85}. 
However, for a given class of topological insulators, one may still distinguish between several subclasses according to further criteria.

In his seminal paper~\cite{Kitaev06}, Kitaev introduced a spin-$1/2$ model defined on the honeycomb lattice that can be mapped onto a free Majorana-fermion problem coupled to a static $\mathbb{Z}_2$ gauge field. According to the tenfold way classification, this system is a class-D topological superconductor characterized by a Chern number $\nu \in \mathbb{Z}$. Importantly, Kitaev has shown that the anyonic properties of the excitations of this model solely depend on  $\nu \mod 16$ giving rise to the celebrated sixteenfold way~\cite{Kitaev06,Bernevig15}. This classification relies on the topological spin  $\theta=\mathrm{e}^{\mathrm{i} \pi \nu/8}$ of vortex excitations that are Abelian (non-Abelian) anyons if $\nu$ is even (odd). 
Experimental signatures of these sixteen different topological orders have been recently proposed for fractional quantum Hall states at half-integer filling factors \cite{Ma19}.

The Chern number $\nu$ counts the number of chiral Majorana edge modes for a system with open boundary conditions. For real fermions, $\nu$ is associated with thermal transport whereas for complex fermions, it is related to electric transport.  To obtain a nonvanishing Chern number, one must break the time-reversal symmetry. As early realized by Kitaev~\cite{Kitaev06}, this symmetry can be broken explicitly (e.g., by adding a magnetic field) or spontaneously (e.g., by adding odd cycles in the lattice \cite{Yao07,Dusuel08_2,Nasu15}). However, an external magnetic field in the Kitaev model leads to interaction terms between Majorana fermions. To break time-reversal symmetry while preserving the integrability of the model, Kitaev suggested to introduce a three-spin term which amounts  to add next-nearest-neighbor hopping terms for the Majorana fermions  on the honeycomb lattice. The corresponding model, which is closely related to the Haldane model for anomalous quantum Hall effect~\cite{Haldane88}, has been the subject of many studies (see, e.g. Refs.~\cite{Lahtinen08,Kamfor10,Lahtinen10,Lahtinen11,Kells11,Lahtinen12,Lahtinen14}), notably at finite temperature \cite{Self17,Self19}.  

The goal of the present paper is to investigate topological phases that can be found in different sectors of the Kitaev honeycomb model in the presence of this time-reversal symmetry-breaking term. More precisely, we analyze the phase diagram of triangular vortex configurations (and their dual) and  show that fourteen (among sixteen) different anyon theories can be generated when varying the strength of  the corresponding coupling term. We also derive a general result about the parity of the Chern number: {\em any periodic vortex configuration with an odd number of vortices per geometric unit cell can only host even Chern numbers whereas odd Chern numbers can be found in other cases}. In addition, we elucidate the origin of gapless phases emerging for a family of vortex configurations lately observed \cite{Zhang20}.

This paper is organized as follows:
In Sec.~\ref{sec:model}, we introduce the model directly in the  Majorana fermion language. Its symmetries are discussed in Sec.~\ref{sec:Symmetries}, and the consequences for the parity of the Chern numbers are detailed in Sec.~\ref{sec:Parity}.  The study of the triangular configurations is presented in Sec.~\ref{sec:Triangular} where a classification in terms of two indicators (vortex-number parity and inversion symmetry) is proposed to analyze the results. Finally, limiting cases and effective models are examined in Sec.~\ref{sec:limiting}.

\newpage

%
%
\section{Model and definitions}
\label{sec:model}
\begin{figure}[t]
\includegraphics[width=0.7\columnwidth]{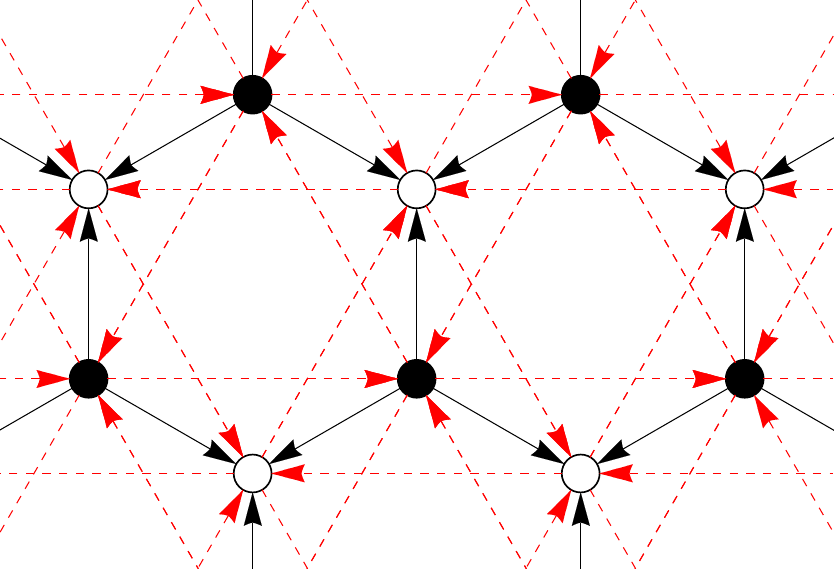}
\caption{ Pictorial representation of the Hamiltonian $H$ for the standard gauge where $u_{jk}=+1$ if $j$ is a white site and $k$ is a black site. This gauge leads to the vortex-free configuration ($w_p=+1$ for all $p$). Black links correspond to nearest-neighbor hoppings $J$  and dashed red links correspond to next-nearest-neighbor hoppings $\kappa$.}
\label{fig:honeycomblattice}
\end{figure}

%
%
We consider the Kitaev honeycomb model in the presence of a three-spin term that breaks time-reversal symmetry~\cite{Kitaev06}. As explained by Kitaev, the original spins $1/2$  defined on the vertices of the honeycomb lattice can be replaced by Majorana operators. This transformation leads to an effective quadratic fermionic Hamiltonian given by
%
%
\be 
H=\frac{\mathrm{i}}{4} \sum_{j,k}   A_{jk} c_j c_k,
\label{eq:ham_Majo}
\ee 
%
%
where the sum is performed over all sites of the honeycomb lattice. The matrix $A$ is a real skew-symmetric matrix whose elements depend on $\mathbb{Z}_2$ link variables \mbox{$u_{jk}=-u_{kj}=\pm 1$} defined on each link of the lattice and where  $c_j$ is a Majorana operator acting on site $j$ (see Ref.~\cite{Kitaev06} for a detailed derivation). Hence, this Hamiltonian describes noninteracting Majorana fermions, coupled to a $\mathbb{Z}_2$ gauge field. For the problem at hand, one has:
\begin{eqnarray}
A_{jk}&=&2 \, J \, u_{jk},  \text{if  $j$ and $k$ are nearest neighbors}, \nonumber\\
A_{jk}&=&2 \, \kappa \, u_{jl} \, u_{lk}, \text{if $j$ and $k$ are next-nearest neighbors}, \nonumber \\
A_{jk}&=&0, \text{otherwise}. \nonumber
\end{eqnarray}
The term proportional to $J$ involves only one gauge variable $u_{jk}$ and readily satisfies $A_{jk}=-A_{kj}$. Without loss of generality, we only consider the case $\kappa \geqslant 0$ and set  $J=1$ in the following. By contrast, the term proportional to $\kappa$ involves the product of two gauge variables $u_{jl} u_{lk}$ (where the site $l$ is connected to sites $j$ and $k$) and requires a specific orientation choice to ensure the skew symmetry of the matrix $A$. Here, we choose $A_{jk}=+2 \, \kappa \, u_{jl} \, u_{lk}$ if the triplet $(k,l,j)$ is oriented clockwise (see Fig.~\ref{fig:honeycomblattice} for the illustration).
 
Following Kitaev, for each plaquette $p$, we define the $\mathbb{Z}_2$ plaquette variable $w_p=\prod_{(j,k) \in p} u_{jk}$ where $j$ belongs to the white sublattice and $k$ belongs to the black sublattice. If \mbox{$w_p=-1$} ($w_p=+1$), we will say that there is a (no) vortex in the plaquette $p$. Two sets of  link variables are said to be equivalent if they lead to the same map of $w_p$'s, i.e., the same vortex configuration. 
%
%
%
%
%
The spectrum of the original spin model is obtained by studying all possible inequivalent configurations of $u_{jk}$'s which define a vortex sector. As a Majorana fermion problem, the spectrum in each sector is symmetric so that the ground-state energy is obtained by summing over all negative-energy levels  (subtleties about the parity of  physical states are discussed in Refs.~\cite{Kitaev06,Pedrocchi11,Zschocke15}). 
In the following, the gap of a given sector refers to the energy difference between the two states around zero energy. 

As shown by Lieb~\cite{Lieb94} in a related problem, the (absolute) ground-state energy at $\kappa=0$ is found in the vortex-free sector where $w_p=+1$ for all plaquettes $p$. For $\kappa \neq 0$, the location of the ground state is still an open question which is not the focus of the present paper.
Let us simply mention that, among all configurations discussed below (triangular vortex lattices and their duals), we find that the ground state is in the vortex-free sector for $\kappa<1.301730(1)$ and in the vortex-full sector otherwise.

%
%
\section{Symmetries}
\label{sec:Symmetries}
%
%
As a (free) Majorana fermion model, $H$ is particle-hole symmetric. In addition, for $\kappa \neq 0$, $H$ breaks time-reversal symmetry~\cite{Kitaev06} and, therefore, is a class-D topological superconductor according to the tenfold classification~\cite{Schnyder08,Kitaev09}. In this section, we discuss the influence of symmetries on the spectrum of the Hamiltonian.

%
\subsection{Translation symmetry}

\begin{figure}[t]
\includegraphics[width=0.8\columnwidth]{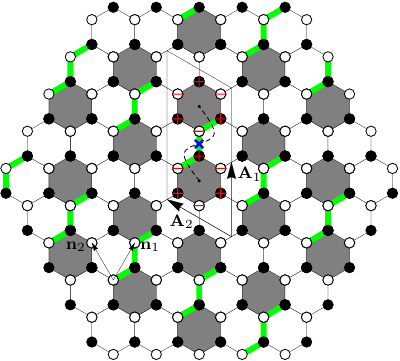}
\caption{Triangular vortex configuration with vortex density $\rho=1/3$. Gray (white) plaquettes indicate plaquettes with (without) a vortex. In this case,  $P_w=-1$, so that the Hamiltonian unit cell (thin rhombus) is twice larger that the geometric unit cell spanned by the vectors $\mathbf{A}_1$ and $\mathbf{A}_2$. The blue cross indicates the inversion center. Black thin (green thick) links correspond to $u_{jk}=+1$ ($u_{jk}=-1$) if $j$ is a white site and $k$ is a black site. The gauge transformation $\mathcal{G}$ is shown as $\pm$ (red symbols) for sites within the Hamiltonian unit cell. The dashed line indicates the string of flipped links between the two gray plaquettes inside the unit cell (see Appendix~\ref{app:gauge1} for a systematic construction).}
\label{fig:1o3}
\end{figure}

In this paper, we will focus on periodic vortex configurations defined by a geometric unit cell spanned by two vectors $\mathbf{A}_1$ and $\mathbf{A}_2$ (see Fig.~\ref{fig:1o3} for illustration).  Denoting by $\mathcal{T}_1$ and $\mathcal{T}_2$, the magnetic translation operators associated with these vectors ($[H,\mathcal{T}_{1}]=[H,\mathcal{T}_{2}]=0$), one has to distinguish between two different cases depending on the parity $P_w=\prod_p w_p$ of the vortex number inside the geometric unit cell.

%
%
%
%
If $P_w=+1$ (even number of vortices per geometric unit cell), then $[ \mathcal{T}_1,\mathcal{T}_2 ]=0$ (see, e.g., Fig.~\ref{fig:2o3}).  
In this case, the energy levels $\varepsilon$ can be labeled by a band index $l$ which runs from 1 to $N$ (number of sites in the geometric unit cell) and a momentum $\mathbf{k}$ that lies into the first Brillouin zone since one has \mbox{$\varepsilon_l(\mathbf{k}+ m \, \mathbf{A}_1^* + n \, \mathbf{A}_2^*)=\varepsilon_l(\mathbf{k})$} where $\mathbf{A}_1^*$ and $\mathbf{A}_2^*$ are the reciprocal lattice vectors and \mbox{$(m,n)\in \mathbb{Z}^2$}. 

By contrast, if $P_w=-1$ (odd number of vortices per geometric unit cell), one has $\{ \mathcal{T}_1,\mathcal{T}_2 \}=0$ and, hence,  $[ \mathcal{T}_1^2,\mathcal{T}_2]= [ \mathcal{T}_1,\mathcal{T}_2^2 ]=0$~\cite{Zhang20} (see, e.g.,  Figs.~\ref{fig:1o3} and \ref{fig:3o4}). This indicates that one must double the unit cell to get commuting translation operators and use Bloch theorem. As a direct consequence, the band index $l$ runs from 1 to $2N$ and one has:
\be
\varepsilon_l(\mathbf{k}+ m \, \mathbf{A}_1^*/2 + n \, \mathbf{A}_2^*/2)=\varepsilon_l(\mathbf{k}).
\label{eq:symm_Bloch}
\ee

%
%
\subsection{Inversion symmetry}
%
%
Some vortex configurations may also have an inversion symmetry with a center that depends on the configuration considered.  The inversion operator $\mathcal{P}=\mathcal{G}\, \mathcal{I}$ is actually composed of a pure spatial inversion $\mathcal{I}$ and a $\mathbb{Z}_2$ gauge transformation $\mathcal{G}$. The pure spatial inversion $\mathcal{I}$ acts as
\be
\mathcal{I}: c_{j} \to c_{-j}, 
\ee
where $\pm j$ stands for the site at position $\pm\mathbf{r}_j$ so that $\mathcal{I}^2=\mathds{1}$. The inversion center stands either  in the middle of a link (as in Fig.~\ref{fig:1o3}) or at the center of a plaquette (as in Figs.~\ref{fig:2o3} and \ref{fig:3o4}). As such, $\mathcal{I}$ exchanges black and white sites. 
The $\mathbb{Z}_2$ gauge transformation acts as
\be
\mathcal{G}: c_{j} \to g_j c_{j}, 
\label{eq:gauge_transfo}
\ee
where $g_j=\pm 1$ are $\mathbb{Z}_2$ site variables that depend on the links variable $u$'s. Again, one has $\mathcal{G}^2=\mathds{1}$. The resulting inversion operator $\mathcal{P}$ is traceless and its square acts as
\be
\mathcal{P}^2: c_{j} \to g_{-j} g_j c_{j}, 
\label{eq:P2}
\ee
so that $\mathcal{P}^2=\pm \mathds{1}$. Indeed, if $[H, \mathcal{P}]=0$, one also has $[H, \mathcal{P}^2]=0$ which implies that the product $g_{-j} g_j$ is independent of $j$ although it can take two values $\pm 1$. If $\mathcal{P}^2=+ \mathds{1}$, the gauge transformation is symmetric under spatial inversion ($[\mathcal{I}, \mathcal{G}]=0$  and $g_{-j}=g_j$). By contrast,  if $\mathcal{P}^2=- \mathds{1}$, the gauge transformation is antisymmetric under spatial inversion ($\{\mathcal{I}, \mathcal{G}\}=0$ and  $g_{-j}=-g_j$).

Finally, if the Hamiltonian has the inversion symmetry, i.e., if $[H,  \mathcal{P}]=0$], one has:
\be
\varepsilon_l(-\mathbf{k})=\varepsilon_l(\mathbf{k}).
\ee
%
%
\subsection{Particle-hole symmetry}
%

The operator $\mathcal{C}$ associated with the particle-hole symmetry is an antiunitary operator which, in the present case, is equivalent to complex conjugation ($\mathcal{C}^2=\mathds{1}$) so that $\{H,\mathcal{C}\}=0$ for all vortex configurations. As a result,
the spectrum has the following symmetry \mbox{$\varepsilon_l(\mathbf{k})\leftrightarrow -\varepsilon_l(-\mathbf{k})$}. 

Interestingly, one has \mbox{$[\mathcal{P},\mathcal{C}]=0$}. Hence, following Zhang {\it et al.}~\cite{Zhang20}, one can define the operator $\mathcal{R}=\mathcal{P}\, \mathcal{C}$ which is a momentum-conserving antiunitary operator such that $\{H,\mathcal{R}\}=0$. As a direct consequence, if $H$ has the inversion symmetry, its spectrum has the following symmetry \mbox{$\varepsilon_l(\mathbf{k}) \leftrightarrow -\varepsilon_l(\mathbf{k})$}.

%
%
\section{Parity of Chern numbers}
\label{sec:Parity}
%
%

As a two-dimensional class-D topological superconductor~\cite{Schnyder08,Kitaev09}, when the system is gapped, it is characterized by a Chern number $\nu \in \mathbb{Z}$. In this section, we establish a criteria to determine the parity of this Chern number for any periodic vortex configuration.

The general considerations on translation symmetries have an important consequence on the parity of the Chern numbers. Indeed, as discussed above, when the number of vortices per geometric unit cell is odd \mbox{($P_w=-1$)}, the spectrum has an extra periodicity in the Brillouin zone stemming from the anticommutation relation of the magnetic translation operators. For clarity, let us assume that the Hamiltonian unit cell is defined by $2 \mathbf{A}_1$ and $\mathbf{A}_2$ (see Fig.~\ref{fig:1o3} for illustration).  

Equation~(\ref{eq:symm_Bloch}) indicates that if the gap closes at a given point $\mathbf{k}_0$ in the first Brillouin zone, then, it also closes at $\mathbf{k}_0+\mathbf{A}_2^*/2$ which is not related to $\mathbf{k}_0$ by a translation of the reciprocal lattice $(\mathbf{A}_1^*/2 , \mathbf{A}_2^*)$. In other words, when $P_w=-1$,  one always has an even number of points in the Brillouin zone where the gap vanishes.
Furthermore, as discussed in Ref.~\cite{Bellissard95}, one can associate with each point where the gap closes a Berry index (which is an integer).  The sum of all Berry indices over all such points gives the variation of the Chern number. In the present context, this variation must be understood as the result of a process where a transition between two gapped phases is obtained by varying some parameters in the Hamiltonian. Thus, since gap-closure points always arise in pairs $(\mathbf{k}_0, \mathbf{k}_0+\mathbf{A}_2^*/2)$ carrying the same Berry index, we straightforwardly obtain that the variation of the Chern number $\Delta \nu$ for configurations with  $P_w=-1$ is always even. 

Up to now, we focused on isotropic couplings ($J$ does not depend on the link orientation). However, if one rather considers anisotropic couplings and the perturbative limit where one of them is much larger than the two others (the isolated-dimer limit), the effective Hamiltonian at $\kappa=0$ is unitarily equivalent to the toric code Hamiltonian~\cite{Kitaev06} (see also Refs.~\cite{Schmidt08,Vidal08_2} for a more detailed discussion). Thus, in this strongly anisotropic limit and for $\kappa=0$, the system is always gapped and time-reversal symmetric so that  $\nu=0$ for all vortex configurations.  Since $\Delta \nu$ is even when one varies the couplings,  we conclude that {\em all gapped vortex configurations with  $P_w=-1$ only host even Chern numbers}. 

By contrast, for  $P_w=+1$, there are no constraints on $\Delta \nu$ so that any Chern number parity can be found in such configurations.

%
\section{Triangular configurations}
\label{sec:Triangular}
%
%

Although class-D topological superconductors are classified by a Chern number $\nu \in \mathbb{Z}$ in two space dimensions, Kitaev has shown that the properties of the anyons in the corresponding topological phase are solely characterized  by $\nu \mod 16$~\cite{Kitaev06}. This celebrated sixteenfold way also indicates that anyons are Abelian if $\nu$ is  even, whereas they are non-Abelian for odd $\nu$. The results derived in Sec.~\ref{sec:Parity} implies that this latter case can only be found in vortex configurations with $P_w=+1$.  In this section, we investigate different vortex configurations in order to exhibit these sixteen different phases. Since changing the sign of $\kappa$ changes the sign of $\nu$~\cite{Kitaev06}, it is actually sufficient to find all possible Chern number $|\nu| \leqslant 8$. 

Any vortex configuration may be of \mbox{interest} in itself and display nontrivial features. Here, we restrict our investigation to triangular vortex configurations and their dual obtained by changing $w_p \rightarrow -w_p$ in each plaquette $p$. By construction, these triangular vortex configurations are spanned by two vectors $\mathbf{A}_1$ and $\mathbf{A}_2$ that are parametrized by two integers $p$ and $q$ such that:
%
%
\begin{eqnarray}
\mathbf{A}_1&=& p \: \mathbf{n}_1+q \: \mathbf{n}_2, \\
\mathbf{A}_2&=&-q \: \mathbf{n}_1+(p+q) \: \mathbf{n}_2,
\end{eqnarray}
%
%
where $\mathbf{n}_1$ and $\mathbf{n}_2$ are the Bravais vectors of the honeycomb lattice (see Fig.~\ref{fig:1o3}). The corresponding vortex density is given by $\rho=1/n$ where $n=p^2+q^2+p \, q$. For dual configurations, the same vectors $\mathbf{A}_1$ and $\mathbf{A}_2$ lead to a vortex density $\rho=(n-1)/n$. Examples are shown in Figs.~\ref{fig:1o3} and \ref{fig:2o3} for $p=1$ and $q=1$.
All these configurations are invariant under inversion symmetry $\mathcal{P}$.

%
%
\subsection{Vortex lattices ($P_w=-1$, $\mathcal{P}^2=-\mathds{1}$)}
\label{subsec:a}
%
%

By construction, all triangular vortex configurations have one vortex per geometric unit cell so that $P_w=-1$. Furthermore, 
these configurations are invariant under inversion $\mathcal{P}=\mathcal{G}\mathcal{I}$ and it is always possible to find a configu\-ration of the link variables and a symmetry center  such that the gauge transformation $\mathcal{G}$ is trivial: $g_j=+1$ if $j$ is a black site and $g_j=-1$ if $j$ is a white site (see Fig.~\ref{fig:1o3} for a concrete example and Appendix \ref{app:gauge1} for a general construction). As can be seen in Eq.~(\ref{eq:P2}), such a transformation corresponds to $\mathcal{P}^2=-\mathds{1}$. 

Since for these configurations, $[H,\mathcal{P}]=0$, one can write  $H=H_{\rm +} \oplus H_{\rm -},$ where the subscript  $\pm$ refers to eigenvalues $\pm \rm{i}$ of $\mathcal{P}$. In this case, the particle-hole symmetry acts as:
$\mathcal{C} H_\pm \mathcal{C} =-H_\mp$ so that
\be 
{\rm spec}(H_\pm)=-{\rm spec}(H_\mp).
\ee
 In other words, the spectrum of $H_+$ and $H_-$ just differ by a sign. In the absence of any additional symmetry, level repulsion makes it unlikely that two eigenvalues of $H$ coincide. We conclude that {\em one cannot get extended gapless phases for vortex configurations with inversion symmetry if  $\mathcal{P}^2=-\mathds{1}$}. In this case, one expects gapped phases separated by gapless points.
 
 For the triangular vortex configurations considered in this section, we computed the Chern number in each gapped phase found in the range $\kappa$ from 0 to $10$. We performed a systematic study of all configurations with  $\rho=1/n$ with $n <30$.
Results are summarized  in Table~\ref{tab:a}.
%
%
\begin{table}[h]
\center
\begin{tabular}{|c |c |c |c |c |c |c |c |c |c|c|c|}
\hline
$1/3$ & $1/4$ & $1/7$ & $1/9$ & $1/12$ & $1/13$ & $1/16$ & $1/19$ & $1/21$ & $1/25$ & $1/27$ & $1/28$\\
\hline
0        & 4        & 2        & 0        & 0         & 2           & 2         & 2           &  0             &  2           &  0         &  2\\
\hline
4        & 0        & -2       & 4        & 4         & -2         & -2         & -2           & 4             & -2           &  4	     &  -2\\
\hline
-2       & 4        & 2        & -2       & 8         & 4          & 4         &             &  -2              & 4          &  -2	     &  4\\
\hline
2        & -8       & -2      & 2         & -4         & 0          & 0       &              &  2               & 0          &  2	     &  0\\
\hline
-2       &           &        & -2        &         & 4          &  4         &             & -2              &  4          &  -2	     &  4\\
\hline
-6       &           &          & 4         &          & -2         &          &             &                       &   -2          &  4	     &  8\\
\hline
          &           &          & 0         &              & 2          &          &             &                    & 2          &  0	     &  -4\\
\hline
         &            &          & 4         &            & -2        &          &             &                        &  -2          &  4	     &  -8\\
\hline
         &            &          & -2        &            &            &         &             &                         & 4          &  -8	     &  4\\
\hline
         &            &          &         &            &            &         &             &                            & 0          &  	     &  \\
\hline
         &            &          &         &            &            &         &             &                            & 4          &  	     &  \\
         \hline
\end{tabular}
\caption{Chern numbers for triangular vortex configurations. The first row indicates the vortex density $\rho=1/n$. Each column gives the set of Chern numbers found in each gapped phases obtained by varying $\kappa$ from 0 to $10$ (from top to bottom). Precise boundaries of these phases are given in  Appendix \ref{app:phaseboundaries}.}
\label{tab:a}
\end{table}
%
%

As anticipated in Sec.~\ref{sec:Symmetries}, since $P_w=-1$ for these configurations, one only finds even Chern numbers. In addition, when $n$ is a multiple of 3, the system is gapped at $\kappa=0$ where the system is time-reversal symmetric~\cite{Kamfor11}, and hence, $\nu=0$. 

Since changing the sign of $\kappa$ changes the sign of $\nu$, these  triangular vortex configurations allow one to generate all possible even $\nu \mod 16$, and hence, all Abelian topological phases expected for free Majorana fermions coupled to a $\mathbb{Z}_2$ gauge field in two dimensions~\cite{Kitaev06}. To generate odd Chern numbers, one must consider configurations with $P_w=+1$. 

%
%
\subsection{Dual vortex lattices with $n$ odd \\
($P_w=+1$, $\mathcal{P}^2=-\mathds{1}$)}
\label{subsec:b}
\begin{figure}[t]
\includegraphics[width=0.8\columnwidth]{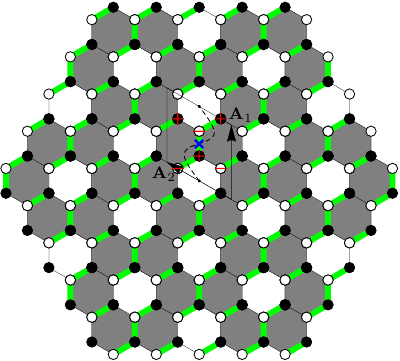}
\caption{Dual vortex configuration $\rho=2/3$ (see Fig.~\ref{fig:1o3} for notations). In this case, $P_w=+1$, so that the Hamiltonian unit cell coincides with the geometric unit cell (see also Appendix~\ref{app:gauge2}).}
\label{fig:2o3}
\end{figure}

%
%
A simple way to generate $P_w=+1$ configurations is to take triangular vortex configurations with $\rho=1/n$ ($n$ odd) and to change $w_p$ into $-w_p$ in each plaquette. This gives rise to a dual vortex lattice with an even number of vortices per geometric unit cell and a vortex density \mbox{$\rho=(n-1)/n$}. 
As previously, we can  systematically find a set of link variables and a symmetry center such that the gauge transformation $\mathcal{G}$ is trivial (see Fig.~\ref{fig:2o3} for a concrete example and  Appendix \ref{app:gauge2} for a general construction) so that, one again finds \mbox{$\mathcal{P}^2=-\mathds{1}$} for this inversion operator. Thus, again, one only gets gapped phases separated by gapless points. 

%
%
%
%

%
%
\begin{table}[h]
\center
\begin{tabular}{|c |c |c |c |c |c |c |c |}
\hline
$2/3$ & $6/7$ & $8/9$  & $12/13$ &  $18/19$ & $20/21$ & $24/25$ & $26/27$  \\
\hline
3               & 2       & 0              & 1                     & 1           &  0         &     1  &     0    \\
\hline
-1             & 1        & 2              & 2                     & 2           & 2         &      2    &      2    \\
\hline
-2             & -2        & 1             & -1                    &  1         &  -1        &     1   &     -1    \\
\hline
                & -5       & -2             & -2                    &  -2            &  -2    &     -2   &     -2  \\
\hline
                & 1      &                   &                         &             &             &           &    \\
\hline
 \end{tabular}
\caption{Chern numbers for dual configurations with vortex densities $\rho=(n-1)/n$ and odd $n$. Conventions are the same as in Table~\ref{tab:a}.}
\label{tab:b}
\end{table}
%
%

Results are summarized in Table \ref{tab:b} where we used the same conventions as Table \ref{tab:a}. Since \mbox{$P_w=+1$}, one can get a Chern number with odd  ($\nu=\pm 1, 3,-5$) or even ($\nu=0,\pm 2$) parity. Contrary to  triangular vortex configurations, these systems are always gapless at $\kappa=0$. Thus, we do not have a simple explanation for these phases with $\nu=0$ emerging at infinitesimal $\kappa>0$ since we cannot connect them to a gapped time-reversal symmetric point. Nevertheless, we observed (up to $n=39$) that such phases are always found for  $n=0 \mod 3$ $(n>3)$. 

%
%
\subsection{Dual vortex lattices with  $n$ even \\
 ($P_w=-1$, $\mathcal{P}^2=+\mathds{1}$)}
 \label{subsec:c}
%
%

%
%
\begin{figure}[t]
\includegraphics[width=0.8\columnwidth]{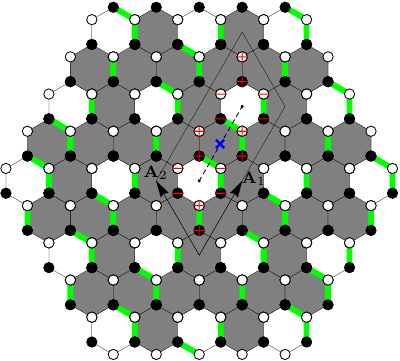}
\caption{Dual vortex configuration $\rho=3/4$ (see Fig.~\ref{fig:1o3} for notations). In this case, $P_w=-1$ so that the Hamiltonian unit cell is twice larger that the geometric unit cell. The dashed line indicates the string of flipped links connecting the two white plaquettes inside the unit cell (see also Appendix~\ref{app:gauge2}).} 
\label{fig:3o4}
\end{figure}
%
%
Let us, now, consider dual vortex configurations obtained from triangular vortex configurations with \mbox{$\rho=1/n$} ($n$ odd) by changing $w_p$ into $-w_p$ in each plaquette. These configurations  have a vortex density $\rho=(n-1)/n$ with $n$ even so that 
$P_w=-1$ (odd number of vortices per geometric unit cell) and, consequently, gapped phases can only host even Chern number. 
However, a salient feature of these configurations is that, contrary to previous cases, the gauge transformation ${\mathcal G}$ involved in the inversion symmetry ${\mathcal P}$ is nontrivial  (see Fig.~\ref{fig:3o4} for a concrete example and Appendix \ref{app:gauge2} for a general construction). As a result, for these configurations one gets $\mathcal{P}^2=+\mathds{1}$.
 
In this case, one can write  \mbox{$H=H_{\rm +} \oplus H_{\rm -}$}, where the subscript  $\pm$ refers to eigenvalues $\pm \rm{1}$ of $\mathcal{P}$. The particle-hole symmetry then acts as:
$\mathcal{C} H_\pm \mathcal{C} =-H_\pm$ so that
\be 
{\rm spec}(H_\pm)=-{\rm spec}(H_\pm).
\ee
 
The spectra of $H_+$ and $H_-$ are not related, and the corresponding energy levels, therefore, have no reason to repel so that overlapping bands are possible. As a consequence, {\em one cannot exclude extended gapless phases for vortex configurations with inversion symmetry if  $\mathcal{P}^2=+\mathds{1}$}.

%
%
\begin{table}[h]
\center
\begin{tabular}{|c |c |c |c |}
\hline
$3/4 $ & $11/12$ & $15/16$ & $27/28$ \\
\hline
-- & 0 & -- & -- \\
\hline
2 & -- &2  & 2 \\
\hline
--& 2 &-- & -- \\
\hline
-2 & -- & -2  & -2 \\
\hline
-- & -2 & --  & --  \\
\hline
-4 & -- & 0  &  \\
\hline
 & 0 & &  \\
 \hline
\end{tabular}
\caption{Chern numbers for dual configurations with vortex densities $\rho=(n-1)/n$ and even $n$. Conventions are the same as in Table~\ref{tab:a}. Gapless phases are denoted by $-$'s.}
\label{tab:c}
\end{table}
%
%

Results for this family of configurations are given in Table~\ref{tab:c}. As expected, since $P_w=-1$, one only gets gapped phases with even Chern number separated by gapless phases. Note also that when $n=0 \mod 12 $, the perturbation due to the triangular lattice of white plaquettes leads to a nesting of the Dirac points in the vortex-full model and opens a gap at $\kappa=0$ (see Ref.~\cite{Kamfor11} for a related discussion). Hence, in this case, one gets $\nu=0$ at $\kappa=0$ due to time-reversal symmetry as can be seen for $\rho=11/12$.  Otherwise, the system remains gapless at small $\kappa$.

%
%
\section{The large-dilution limits}
\label{sec:limiting}
%
%
A close inspection of Tables~\ref{tab:a}-\ref{tab:c} unveils several features in the large-$n$ limit. In this section, we give some arguments to understand the Chern numbers found in this limit.  
%
%
\subsection{Around the vortex-free background}
%
%
%
%
\begin{figure}[t]
\includegraphics[width=0.65\columnwidth]{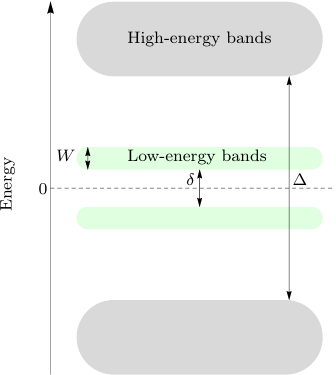}
\caption{Schematic energy spectrum of  dilute configurations. When gray (white) plaquettes are strongly dilute in a vortex-free (vortex-full) background, the bandwidth of the low-energy bands $W$ vanishes, $\Delta$ is given by the gap of the vortex-free (vortex-full) sector, and $\delta$ vanishes (remains finite).}
\label{fig:spectrum}
\end{figure}
%
%

The limit of sparse vortex configurations has been discussed by Lahtinen and co-workers  in Refs.~\cite{Lahtinen12, Lahtinen14} who derived an effective model to describe the spectrum and the Chern number of such configurations (see also Ref.~\cite{Grosfeld06} for a related study). In the vortex-free configuration, plaquette excitations consists in pairs of plaquettes with $w_p=-1$ (gray plaquettes, also known as, vortices). Since, in this sector, $\nu$ is odd,  there is 
a single unpaired Majorana zero mode bound to each vortex inside the bulk gap~\cite{Kitaev06}. Thus, in the presence of a dilute vortex lattice in a vortex-free background, the spectrum can be split into high-energy bands (coming from the vortex-free background) and low-energy bands  (coming from tunneling of Majorana fermions between zero modes attached to vortices). The total Chern number $\nu=\nu_h+\nu_l$  is then simply the sum of the high-energy bands Chern number $\nu_h=1$ and the low-energy bands Chern number $\nu_l$. 
For triangular vortex configurations, including only the most relevant tunneling terms, this effective model predicts $\nu=1\pm1$ or $1\pm3$ \cite{Lahtinen12, Lahtinen14} which is in agreement with our findings for density $\rho=1/n$ at large $n$ (see Table~\ref{tab:a}). 

Of course, this effective model is only valid if the total band width of the low-energy bands (which is essentially given by the tunneling terms) is much smaller than the gap between high-energy bands.  Using notations of Fig.~{\ref{fig:spectrum}}, this corresponds to the regime  $\delta+2 W \ll \Delta$. However, the dependence of the tunneling terms on $\kappa$ and $n$ is nontrivial, and each case must be considered carefully to determine the validity range of this approach. Finally, any infinitesimal tunneling term opens a gap at zero energy so that the strongly dilute limit is nontrivial as can be seen in Table~\ref{tab:a}.

%
%
\subsection{Around the vortex-full background}
%
%

\begin{figure}[t]
\includegraphics[width=\columnwidth]{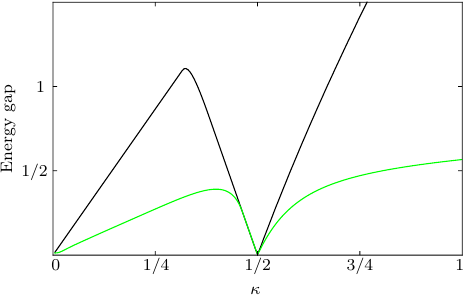} 
\caption{Energy gaps $\Delta$ (black) of the vortex-full configuration (see Appendix~\ref{app:gaps} for an analytical expression) and $\delta$ (green) of isolated white plaquettes in a vortex-full  background as a function of $\kappa$. Both gaps vanish for $\kappa=0,1/2$ whereas \mbox{$\Delta \sim 2 \sqrt{3} \: \kappa$} and $\delta=0.7865(1)$ in the large-$\kappa$ limit.}
\label{fig:splitting}
\end{figure}

Following the study of Lahtinen {\it et al.}~\cite{Lahtinen12, Lahtinen14} around the vortex-free lattice,  let us now consider the vortex-full lattice where $w_p=-1$ for all plaquettes. This vortex configuration ($\rho=1$) has been studied in Refs.~\cite{Lahtinen08,Lahtinen10} in the small-$\kappa$ limit where a gapped phase with $\nu=2$ has been found.  However, as shown in Fig.~\ref{fig:splitting}, there exists a phase transition separating a phase with $\nu=2$ for $0<\kappa<1/2$ from a phase at $\nu=-2$ for $\kappa>1/2$. In the vortex-full configuration, elementary plaquette excitations consist in pairs of  plaquettes with $w_p=+1$ (white plaquettes). Since, in this sector, $\nu$ is even, a white plaquette does not bind an unpaired Majorana mode at zero energy~\cite{Kitaev06} but rather several modes at finite energy~\cite{Volovik99}. In the presence of a dilute white-plaquettes lattice in a vortex-full background, the spectrum splits into high-energy bands (coming from the vortex-full background) and low-energy bands  (coming from the coupling between modes bound to white plaquettes) as depicted in Fig.~\ref{fig:spectrum}. Thus, again, the total Chern number is the sum of high- and low-energy bands Chern numbers, i.e., $\nu=\nu_h+\nu_l$.
Interestingly, when white plaquettes are sufficiently far from each other (large-dilution limit), the tunneling terms between modes bound to these plaquettes vanish so that the low-energy bandwidth $W$ goes to zero. However, on each white plaquette, the two lowest-energy bound modes have an energy $\pm \delta/2$, where $\delta$ depends on $\kappa$ (see Fig.~\ref{fig:splitting}). The corresponding low-energy (flat) bands are topologically trivial ($\nu_l=0$) since they correspond to an ``atomic" limit. Hence, provided $\delta$ is smaller than the bulk gap $\Delta$, we expect the total Chern number  to be $\nu=\nu_h$ for all sparse white-plaquette configurations. Thus, in the large-dilution limit, the phase diagram is the same as the one of the vortex-full sector.
%
%
%
%

As can be seen in Tables~\ref{tab:b} and \ref{tab:c}, one, indeed, gets $\nu=2$ for $0\lesssim \kappa\lesssim1/2$ and $\nu=-2$ for $1/2\lesssim \kappa$ (see Appendix \ref{app:phaseboundaries} for phase boundaries). Nevertheless, for finite but large $n$, one observes the existence of intermediate gapped phases (with $\nu=0,\pm1$) or gapless phases near the points $\kappa=0,1/2$ where both  $\delta$ and  $\Delta$ go to zero (see Fig.~\ref{fig:splitting}). The extension of these intermediate phases shrinks to zero when $n$ increases. In addition, one also observes in Table~\ref{tab:c} that the phase at $\nu=-2$ found for $\kappa>1/2$ ends up at a value of $\kappa$ which increases with $n$ (see Appendix \ref{app:phaseboundaries}) giving rise to a gapless phase and possibly other phases for larger $\kappa$. 

\section{Outlook}
\label{sec:oulook}
%
%

We have shown that a wide variety of topological phases characterized by the Chern number $\nu \mod 16$ could be produced in the Kitaev honeycomb model in the presence of  a simple time-reversal breaking term~\cite{Kitaev06}. In the triangular vortex configurations (and their dual) studied in this paper, we found all possible values except $\nu=\pm 7$. Although, only a finite number of examples were considered, we believe that $\nu=\pm 7$ does not exist for these families of configurations. Indeed, we proved that odd Chern numbers can only be found when the geometric unit cell contains an even number of vortices ($P_w=1$). For the configurations considered here, this concerns dual triangular lattices with vortex density $\rho=(n-1)/n$ and odd $n$.  As can be seen in Table \ref{tab:b}, $\nu=\pm 7$ is absent for $n<30$. For larger $n$, we argued that one simply recovers the vortex-full phase diagram ($\nu=\pm 2$)  except in narrow regions near $\kappa=0$ and $\kappa=1/2$ where one only gets $\nu=0,\pm1$. Nevertheless, the missing Chern numbers $\nu=\pm 7$ are certainly present in other vortex sectors with $P_w=1$.

In this paper, we investigated inversion-symmetric periodic vortex lattices and found that they are classified by two indicators: The parity of the number of vortices per geometric unit cell $P_w=\pm 1$ and the square of the inversion operator  $\mathcal{P}^2=\pm \mathds{1}$. Here, we found only three classes: $(P_w,\mathcal{P}^2)=(-1,-\mathds{1})$ (see Sec.~\ref{subsec:a}), $(+1,-\mathds{1})$ (see Sec.~\ref{subsec:b}), and $(-1,+\mathds{1})$ (see Sec.~\ref{subsec:c}). For other inversion-symmetric vortex lattices, one may create the fourth class $(P_w,\mathcal{P}^2)=(+1,+\mathds{1})$, which should exhibit even and odd Chern numbers as well as gapless phases.

Finally, it would be worth developing an effective model for dilute lattices of white plaquettes in a vortex-full background following the general strategy of Refs.~\cite{Lahtinen12,Lahtinen14}. A main difference is that low-energy bands are built from finite-energy mid-gap states bound to white plaquettes instead of unpaired Majorana zero modes bound to vortices. Such an effective model should help us to understand how the infinitely dilute white-plaquette limit nucleates non-trivial phases as the density of white plaquettes is increased. 

To conclude, we hope that the present paper will motivate further studies to complete the quest of the sixteen topological orders in the Kitaev honeycomb model recently initiated by Zhang {\it et al.} \cite{Zhang20}.

\medskip

\acknowledgments
We thank A. Auerbach, B. A. Bernevig, B. Dou\c{c}ot, S.~Iblisdir, A. Mesaros,  F. Pi\'echon, and G. J. Sreejith  for useful discussions. S. P. was supported by an Erasmus+ International Credit Mobility Grant.

\appendix

%
%
\section{Systematic construction of the $\mathbb{Z}_2$ gauge field and of the gauge transformation $\mathcal{G}$}
\label{app:gauge}
%
%

In this appendix, we propose a systematic construction of the $\mathbb{Z}_2$ gauge field (i.e., the $u_{jk}$'s defining the vortex configuration) with the minimal periodicity for the triangular vortex lattices (and their duals) as well as the gauge transformation $\mathcal{G}$ [i.e., the $g_{j}$'s defined in Eq.~(\ref{eq:gauge_transfo})] involved in the inversion operator $\mathcal{P}=\mathcal{G}\mathcal{I}$. In  both cases, we start from the standard gauge of the vortex-free configuration depicted in Fig.~\ref{fig:honeycomblattice}.  With respect to this standard gauge, flipping a link ($u_{jk} \rightarrow - u_{jk}$) is depicted graphically as a thick green link whereas a thin black link refers to an unflipped one.

%
%
\begin{figure}[t]
\includegraphics[width=0.8\columnwidth]{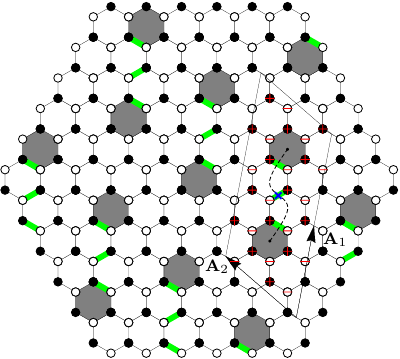}\\
\medskip
\includegraphics[width=0.8\columnwidth]{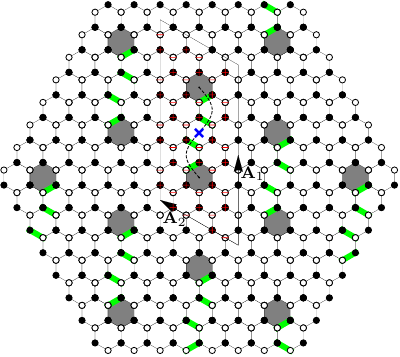}
\caption{Triangular vortex configurations (\mbox{$P_w=-1$}) with \mbox{$\rho=1/7$} (top) and \mbox{$\rho=1/12$} (bottom). The dashed line indicates the string of flipped links (green) between two vortices.  The inversion center lies in the middle of a link for $\rho=1/7$ and in the center of a white plaquette for $\rho=1/12$.  In both cases, the gauge transform $\mathcal{G}$  ($\pm$ red symbols) is antisymmetric ($g_{-j}=-g_j$) so that \mbox{$\mathcal{P}^2=-\mathds{1}$}.}
\label{fig:1o7_1o12}
\end{figure}
%

%
%
\subsection{Triangular vortex lattices}
\label{app:gauge1}
%
%
For the triangular vortex lattices with vortex density $\rho=1/n$ $(P_w=-1)$, there is a systematic way to build a  $\mathbb{Z}_2$ gauge field which is symmetric with respect to the inversion center. First, one has to select an inversion center which, for odd (even) $n$, can always be chosen as the middle of a link (white plaquette). Then, one creates a pair of vortices that one moves until they are in a relative position given by $\mathbf{A}_1$. This is easily achieved by considering a string of flipped links chosen  symmetrically with respect to the inversion center (see Figs.~\ref{fig:1o3} and \ref{fig:1o7_1o12} for examples). Finally, we repeat this procedure periodically with the translation vectors $2 \mathbf{A}_1$ and $\mathbf{A}_2$. 

By construction, the resulting $\mathbb{Z}_2$ gauge field is symmetric under inversion. Hence, the gauge transform $\mathcal{G}$ is trivial: \mbox{$g_j=+1$ $(-1)$} if $j$ is a black (white) site and has the periodicity of the honeycomb. This transformation $\mathcal{G}$ is antisymmetric under inversion, ($g_{-j}=-g_j$), so that $\mathcal{P}^2=-\mathds{1}$.

%
\begin{figure}[t]
\includegraphics[width=0.8\columnwidth]{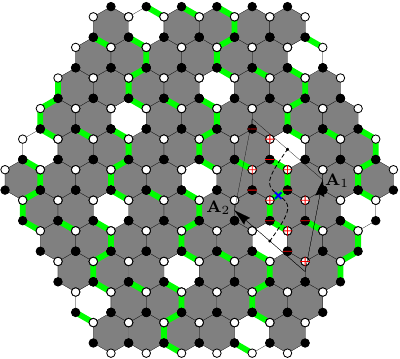}\\
\medskip
\includegraphics[width=0.8\columnwidth]{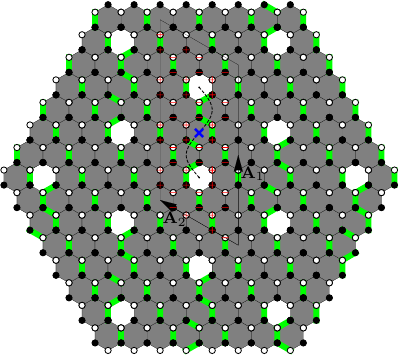}

\caption{Dual vortex configurations with $\rho=6/7$ (top) and $\rho=11/12$ (bottom). The dashed line indicates the string of flipped links (green) between two white plaquettes. For $\rho=6/7$ ($P_w=+1$), the gauge transform $\mathcal{G}$  ($\pm$ red symbols) is antisymmetric ($g_{-j}=-g_j$) so that \mbox{$\mathcal{P}^2=-\mathds{1}$}. By contrast, for $\rho=11/12$ ($P_w=-1$), the gauge transform $\mathcal{G}$ is symmetric ($g_{-j}=g_j$) so that \mbox{$\mathcal{P}^2=+\mathds{1}$}.
}
\label{fig:6o7_11o12}
\end{figure}
%
%

%

%
%
\subsection{Dual lattices}
\label{app:gauge2}
%
%

To build the dual lattice with vortex density \mbox{$\rho=(n-1)/n$}, we start from the gauge used for the triangular lattice configuration with $\rho=1/n$, and we add the missing links to change $w_p$ into $-w_p$ for all plaquettes. This last step can always be achieved by respecting the Hamiltonian unit cell periodicity. However, one has to distinguish between two cases according to the parity of $n$. 

Indeed, when $n$ is odd ($P_w=+1$), this completion step can be performed symmetrically with respect to the inversion center which is the middle of a link [see Figs.~\ref{fig:2o3} and \ref{fig:6o7_11o12} (top) for examples]. Thus, in this case, the resulting $\mathbb{Z}_2$ gauge field is symmetric under inversion; the gauge transform $\mathcal{G}$ is trivial: \mbox{$g_j=+1$ $(-1)$} if $j$ is a black (white) site; $\mathcal{G}$ is antisymmetric under inversion, ($g_{-j}=-g_j$) so that $\mathcal{P}^2=-\mathds{1}$. 

By contrast, when $n$ is even  ($P_w=-1$), one cannot add the missing links symmetrically because the inversion center is the center of a gray plaquette which, as such, must contain an odd number of flipped links  [see Figs.~\ref{fig:3o4} and \ref{fig:6o7_11o12} (bottom) for examples]. The corresponding gauge transform $\mathcal{G}$ is less trivial than previously and can be built according to the following rules: If inversion maps a flipped (unflipped)  link $(j,k)$ onto a flipped (unflipped) link then $g_j=-g_k$. By contrast, if a flipped (unflipped) link $(j,k)$ is mapped onto an unflipped (flipped) link, then, $g_j=g_k$. Such a gauge transform $\mathcal{G}$ is symmetric under inversion ($g_{-j}=g_j$) and one has $\mathcal{P}^2=+\mathds{1}$.

%
%
\section{Phase boundaries of triangular vortex lattices and their dual}
\label{app:phaseboundaries}
%
%
In this appendix, we give the boundaries of each phase obtained for all triangular vortex configurations and their dual discussed in the main text. Changing the relative sign of $\kappa/J$ simply changes the sign of the Chern number. Hence, without loss of generality, we set $J=1$ and only consider the case of $\kappa \geqslant 0$. Furthermore, for simplicity, we restrict our discussion to the range \mbox{$\kappa \in [0,10]$} and give only a relative precision of $10^{-6}$ on the phase boundaries. 
For each integers $p$ and $q$ considered, we show two figures representing a piece of infinite lattice for the triangular vortex configuration with vortex density $\rho=1/n$ (left) and its dual corresponding to $\rho=(n-1)/n$ (right) where $n=p^2+q^2+p\,q$. We denote each Chern number by $\nu_\rho^i$ where the superscript $i$ labels each phase. 

Finally, in many cases, we also found other phases at larger $\kappa$  (e.g.,  $\nu=-14$ for  $\rho=1/7$ and $\kappa>16.4268$), but none of them has a Chern number $\pm 7$ which is the missing one in our list to complete the sixteenfold classification~\cite{Kitaev06}.

\newpage
%
%
\subsection{$(p,q)=(1,0)$}
%
%

%
%
\begin{figure}[h]
\includegraphics[width=0.3\columnwidth]{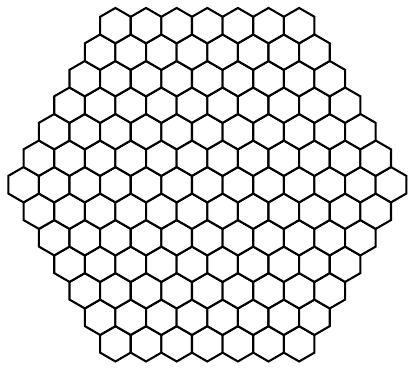} \hspace{20mm}
\includegraphics[width=0.3\columnwidth]{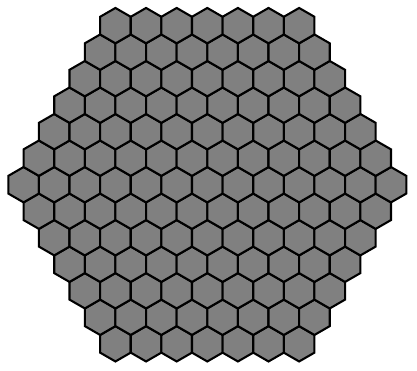} \\
\label{fig:rho_0_1}
\end{figure}
%
%

$\bullet$ $\rho=0$:
%
%
\begin{eqnarray}
\nu_{0}^a&=&+1 \:\: {\rm for}  \:\: 0<\kappa. \nonumber
\label{eq:rho_rho_0}
\end{eqnarray}
%
%

$\bullet$ $\rho=1$:
%
%
\begin{eqnarray}
\nu_{1}^a&=&+2 \:\: {\rm for}  \:\: 0<\kappa<1/2, \nonumber \\
\nu_{1}^b&=&-2  \:\: {\rm for}  \:\: 1/2<\kappa.\nonumber
\label{eq:chern_rho_1}
\end{eqnarray}
%
%

%
%
\subsection{$(p,q)=(1,1)$}
%
%

%
%
\begin{figure}[h]
\includegraphics[width=0.3\columnwidth]{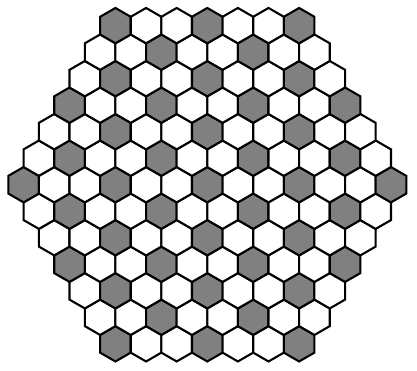} \hspace{20mm}
\includegraphics[width=0.3\columnwidth]{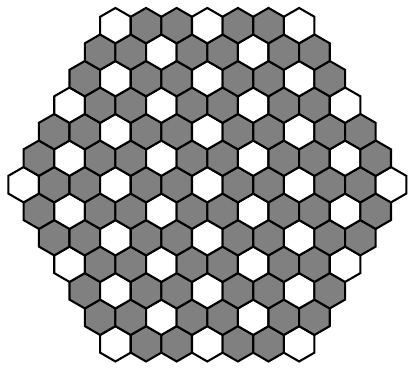} \\
\label{fig:rho_1_1}
\end{figure}
%
%

$\bullet$ $\rho=1/3$:
%
%
\begin{eqnarray}
\nu_{1/3}^a&=&0 \:\: {\rm for}  \:\: 0\leqslant\kappa<0.224745, \nonumber\\
\nu_{1/3}^b&=&+4 \:\: {\rm for}  \:\: 0.224745<\kappa<0.242614, \nonumber\\
\nu_{1/3}^c&=&-2  \:\: {\rm for}  \:\: 0.242614<\kappa<0.313859, \nonumber\\
  \nu_{1/3}^d&=&+2  \:\: {\rm for}  \:\: 0.313859<\kappa<2.22474, \nonumber\\
\nu_{1/3}^e&=&-2  \:\: {\rm for}  \:\: 2.22474<\kappa<3.18614, \nonumber\\
\nu_{1/3}^f&=&-6  \:\: {\rm for}  \:\: 3.18614<\kappa \leqslant 10.\nonumber
\label{eq:chern_1_over_3}
\end{eqnarray}
%
%

$\bullet$ $\rho=2/3$: 
%
%
\begin{eqnarray}
\nu_{2/3}^a&=&+3 \:\: {\rm for}  \:\: 0<\kappa<0.288675, \nonumber\\
\nu_{2/3}^b&=&-1  \:\: {\rm for}  \:\: 0.288675<\kappa<1.1547, \nonumber\\
\nu_{2/3}^c&=&-2  \:\: {\rm for}  \:\: 1.1547<\kappa \leqslant 10.\nonumber
\label{eq:chern_2_over_3}
\end{eqnarray}
%

\newpage

%
%
\subsection{$(p,q)=(2,0)$}
%
%

%
%
\begin{figure}[h]
\includegraphics[width=0.3\columnwidth]{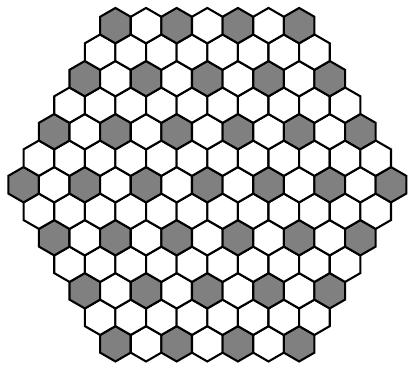} \hspace{20mm}
\includegraphics[width=0.3\columnwidth]{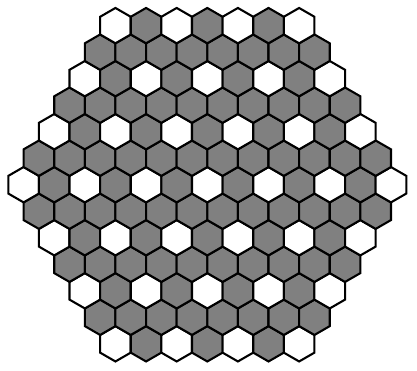}\\
\label{fig:rho_2_0}
\end{figure}
%
%

$\bullet$ $\rho=1/4$:
%
%
\begin{eqnarray}
\nu_{1/4}^a&=&+4 \:\: {\rm for}  \:\: 0<\kappa<0.144327, \nonumber\\
\nu_{1/4}^b&=&0  \:\: {\rm for}  \:\: 0.144327<\kappa<2.46291 , \nonumber\\
\nu_{1/4}^c&=&+4  \:\: {\rm for}  \:\: 2.46291<\kappa<2.54644, \nonumber\\
\nu_{1/4}^c&=&-8  \:\: {\rm for}  \:\: 2.54644<\kappa \leqslant 10.\nonumber
\label{eq:chern_1_over_4}
\end{eqnarray}
%
%

$\bullet$ $\rho=3/4$:
%
%
\begin{eqnarray}
&\text{gapless} &  \:\: {\rm for}  \:\: 0<\kappa<0.30819, \nonumber\\
\nu_{3/4}^a&=&+2 \:\: {\rm for}  \:\: 0.30819<\kappa<0.353553, \nonumber\\
&\text{gapless}&  \:\: {\rm for}  \:\: 0.353553<\kappa<1.04121, \nonumber\\
\nu_{3/4}^b&=&-2  \:\: {\rm for}  \:\: 1.04121<\kappa<1.07809, \nonumber\\
&\text{gapless}&  \:\: {\rm for}  \:\: 1.07809<\kappa<1.37117, \nonumber\\
\nu_{3/4}^c&=&-4  \:\: {\rm for}  \:\: 1.37117 <\kappa \leqslant 10. \nonumber
\label{eq:chern_1_over_4}
\end{eqnarray}
%
%


%
%
\subsection{$(p,q)=(2,1)$}
%
%

%
\begin{figure}[h]
\includegraphics[width=0.3\columnwidth]{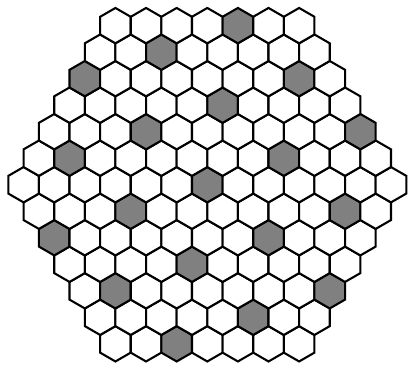} \hspace{20mm}
\includegraphics[width=0.3\columnwidth]{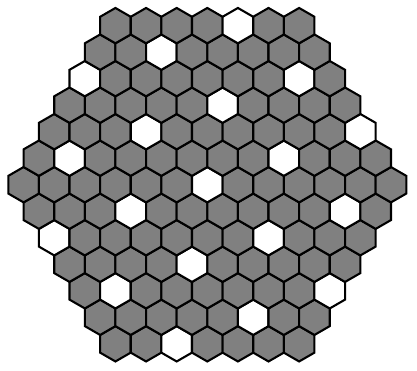}\\
\label{fig:rho_2_1}
\end{figure}
%
%

$\bullet$ $\rho=1/7$:
%
%
\begin{eqnarray}
\nu_{1/7}^a&=& +2 \:\: {\rm for}  \:\: 0<\kappa<0.734537, \nonumber \\
\nu_{1/7}^b&=& -2 \:\: {\rm for}  \:\: 0.734537<\kappa<2.28926, \nonumber \\
\nu_{1/7}^c&=& +2 \:\: {\rm for}  \:\: 2.28926<\kappa<8.02591, \nonumber \\
\nu_{1/7}^d&=& -2 \:\: {\rm for}  \:\: 8.02591<\kappa \leqslant 10. \nonumber
\label{eq:chern_1_over_7}
\end{eqnarray}
%
%

$\bullet$ $\rho=6/7$:
%
%
\begin{eqnarray}
\nu_{6/7}^a&=& +2 \:\: {\rm for}  \:\: 0<\kappa<0.494872, \nonumber\\
\nu_{6/7}^b&=& +1 \:\: {\rm for}  \:\: 0.494872<\kappa<0.518436, \nonumber\\
\nu_{6/7}^c&=& -2 \:\: {\rm for}  \:\: 0.518436<\kappa<0.816497, \nonumber\\
\nu_{6/7}^e&=& -5 \:\: {\rm for}  \:\: 0.816497<\kappa<1.80057, \nonumber\\
\nu_{6/7}^e&=& +1 \:\: {\rm for}  \:\: 1.80057<\kappa \leqslant 10.\nonumber
\label{eq:chern_6_over_7}
\end{eqnarray}
%
%
\newpage

%
%
\subsection{$(p,q)=(3,0)$}
%
%

%
\begin{figure}[h]
\includegraphics[width=0.3\columnwidth]{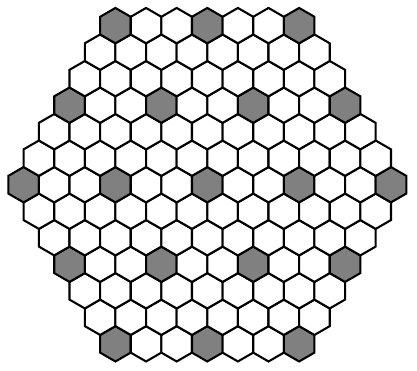} \hspace{20mm}
\includegraphics[width=0.3\columnwidth]{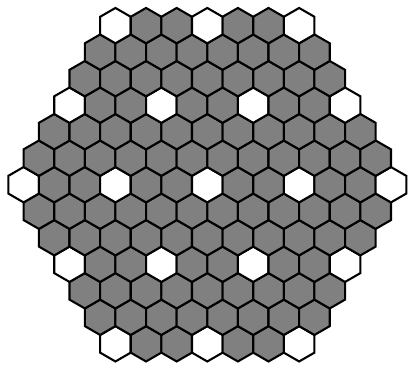}\\
\label{fig:rho_3_0}
\end{figure}
%

$\bullet$ $\rho=1/9$:
%
%
\begin{eqnarray}
\nu_{1/9}^a&=& 0 \:\: {\rm for}  \:\: 0\leqslant\kappa<0.152417, \nonumber\\
\nu_{1/9}^b&=& +4 \:\: {\rm for}  \:\: 0.152417<\kappa<0.16673, \nonumber\\
\nu_{1/9}^c&=& -2 \:\: {\rm for}  \:\: 0.16673<\kappa<0.175741,\nonumber \\
\nu_{1/9}^d&=& +2 \:\: {\rm for}  \:\: 0.175741<\kappa<0.455118, \nonumber\\
\nu_{1/9}^e&=& -2 \:\: {\rm for}  \:\: 0.455118<\kappa<0.656106, \nonumber\\
\nu_{1/9}^f&=& +4 \:\: {\rm for}  \:\: 0.656106<\kappa<0.826569,\nonumber \\
\nu_{1/9}^g&=&  0 \:\: {\rm for}  \:\: 0.826569<\kappa<4.2224 , \nonumber\\
\nu_{1/9}^h&=&  +4 \:\: {\rm for}  \:\: 4.2224<\kappa<4.95638, \nonumber\\
\nu_{1/9}^i&=&  -2 \:\: {\rm for}  \:\: 4.95638<\kappa \leqslant 10.\nonumber
\label{eq:chern_1_over_9}
\end{eqnarray}
%
%

$\bullet$ $\rho=8/9$:
%
%
\begin{eqnarray}
\nu_{8/9}^a&=& 0 \:\: {\rm for}  \:\: 0<\kappa<0.215646, \nonumber\\
\nu_{8/9}^b&=& +2 \:\: {\rm for}  \:\: 0.215646<\kappa<0.445821, \nonumber\\
\nu_{8/9}^c&=& +1 \:\: {\rm for}  \:\: 0.445821<\kappa<0.544229, \nonumber\\
\nu_{8/9}^e&=& -2  \:\: {\rm for}  \:\: 0.544229<\kappa \leqslant 10.\nonumber
\label{eq:chern_1_over_9}
\end{eqnarray}
%

\newpage

%
%
\subsection{$(p,q)=(2,2)$}
%
%

%
%
\begin{figure}[h]
\includegraphics[width=0.3\columnwidth]{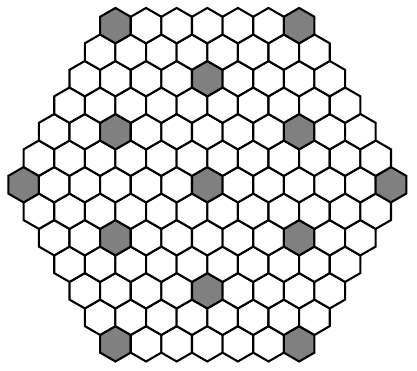} \hspace{20mm}
\includegraphics[width=0.3\columnwidth]{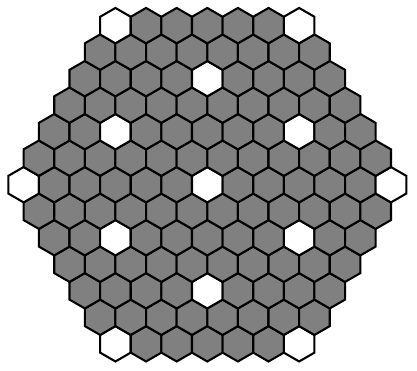}\\
\label{fig:rho_2_2}
\end{figure}
%

$\bullet$ $\rho=1/12$:
%
%
\begin{eqnarray}
\nu_{1/12}^a&=& 0 \:\: {\rm for}  \:\: 0\leqslant\kappa<1.58468,\nonumber \\
\nu_{1/12}^b&=&  +4 \:\: {\rm for}  \:\: 1.58468<\kappa<3.07082,\nonumber \\
\nu_{1/12}^c&=&  +8 \:\: {\rm for}  \:\: 3.07082<\kappa< 9.08762, \nonumber\\
\nu_{1/12}^d&=&   -4 \:\: {\rm for}  \:\: 9.08762<\kappa \leqslant 10.\nonumber
\label{eq:chern_1_over_12}
\end{eqnarray}
%
%

$\bullet$ $\rho=11/12$:
%
%
\begin{eqnarray}
\nu_{11/12}^a&=&0 \:\: {\rm for}  \:\: 0\leqslant\kappa<0.145127, \nonumber\\
&\text{gapless}&  \:\: {\rm for}  \:\: 0.145127<\kappa<0.197929, \nonumber\\
\nu_{11/12}^b&=&+2  \:\: {\rm for}  \:\: 0.197929<\kappa<0.484904, \nonumber\\
&\text{gapless}&  \:\: {\rm for}  \:\: 0.484904<\kappa<0.557619,\nonumber \\
\nu_{11/12}^c&=& -2  \:\: {\rm for}  \:\: 0.557619 <\kappa<2.68462, \nonumber\\
&\text{gapless}&  \:\: {\rm for}  \:\: 2.68462<\kappa<4.6313,\nonumber \\
\nu_{11/12}^d&=& 0  \:\: {\rm for}  \:\:4.6313 <\kappa \leqslant 10.\nonumber
\label{eq:chern_11_over_12}
\end{eqnarray}
%
%


%
%
\subsection{$(p,q)=(3,1)$}
%
%

%
%
\begin{figure}[h]
\includegraphics[width=0.3\columnwidth]{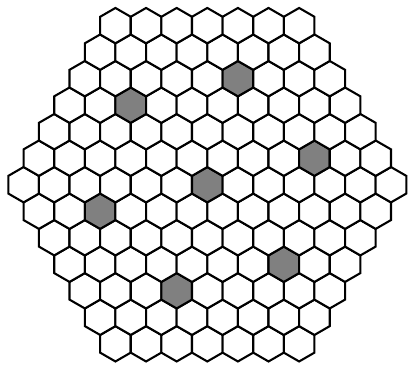} \hspace{20mm}
\includegraphics[width=0.3\columnwidth]{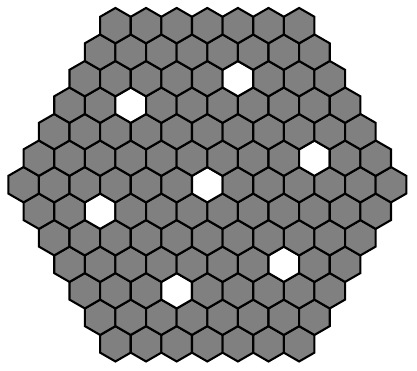}\\
\label{fig:rho_3_1}
\end{figure}
%

$\bullet$ {\em $\rho=1/13$}:
%
%
\begin{eqnarray}
\nu_{1/13}^a&=& +2 \:\: {\rm for}  \:\: 0<\kappa<0.0686732,\nonumber \\
\nu_{1/13}^b&=&  -2 \:\: {\rm for}  \:\: 0.0686732<\kappa<0.0967216,\nonumber \\
\nu_{1/13}^c&=&  +4 \:\: {\rm for}  \:\: 0.0967216<\kappa< 0.111067,\nonumber \\
\nu_{1/13}^d&=&   0 \:\: {\rm for}  \:\: 0.111067<\kappa<0.268735, \nonumber\\
\nu_{1/13}^e&=&  4 \:\: {\rm for}  \:\: 0.268735<\kappa<0.275962, \nonumber\\
\nu_{1/13}^f&=&  -2 \:\: {\rm for}  \:\: 0.275962<\kappa<0.290525, \nonumber\\
\nu_{1/13}^g&=&  +2 \:\: {\rm for}  \:\: 0.290525<\kappa<2.26874, \nonumber\\
\nu_{1/13}^h&=&  -2 \:\: {\rm for}  \:\: 2.26874<\kappa \leqslant 10.\nonumber
\label{eq:chern_1_over_13}
\end{eqnarray}
%
%

$\bullet$ {\em $\rho=12/13$}:
%
%
\begin{eqnarray}
\nu_{12/13}^a&=&+1 \:\: {\rm for}  \:\: 0<\kappa<0.219461, \nonumber\\
\nu_{12/13}^b&=&+2  \:\: {\rm for}  \:\:  0.219461<\kappa<0.505106, \nonumber\\
\nu_{12/13}^c&=& -1  \:\: {\rm for}  \:\: 0.505106<\kappa<0.541875, \nonumber\\
\nu_{12/13}^d&=& -2  \:\: {\rm for}  \:\:0.541875 <\kappa \leqslant 10.\nonumber 
\label{eq:chern_12_over_13}
\end{eqnarray}
%
%

%
%
\subsection{$(p,q)=(4,0)$}
%
%

%
%
\begin{figure}[h]
\includegraphics[width=0.3\columnwidth]{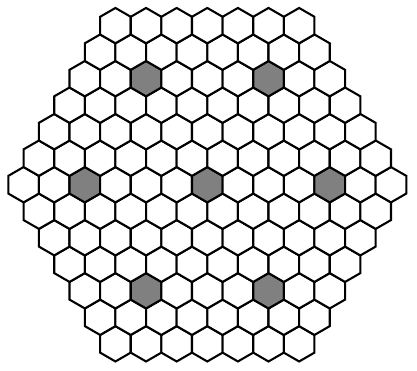} \hspace{20mm}
\includegraphics[width=0.3\columnwidth]{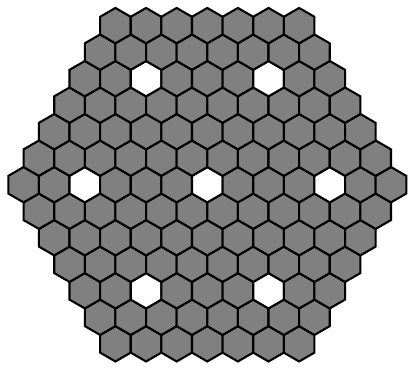}\\
\label{fig:rho_4_0}
\end{figure}
%

$\bullet$ {\em $\rho=1/16$}:
%
%
\begin{eqnarray}
\nu_{1/16}^a&=& +2 \:\: {\rm for}  \:\: 0<\kappa<0.166568, \nonumber\\
\nu_{1/16}^b&=&  -2 \:\: {\rm for}  \:\: 0.166568<\kappa<0.170668, \nonumber\\
\nu_{1/16}^c&=&  +4 \:\: {\rm for}  \:\: 0.170668<\kappa< 0.179436, \nonumber\\
\nu_{1/16}^d&=&   0 \:\: {\rm for}  \:\: 0.179436<\kappa<4.63642, \nonumber\\
\nu_{1/16}^e&=&  +4 \:\: {\rm for}  \:\: 4.63642<\kappa \leqslant 10.\nonumber
\label{eq:chern_1_over_16}
\end{eqnarray}
%
%

$\bullet$ {\em $\rho=15/16$}:

%
%
\begin{eqnarray}
&\text{gapless} &  \:\: {\rm for}  \:\: 0<\kappa<0.111802, \nonumber\\
\nu_{15/16}^a&=&+2 \:\: {\rm for}  \:\: 0.111802<\kappa<0.488697,\nonumber \\
&\text{gapless}&  \:\: {\rm for}  \:\: 0.488697<\kappa<0.540224, \nonumber\\
\nu_{15/16}^b&=&-2  \:\: {\rm for}  \:\: 0.54022<\kappa<8.26214, \nonumber\\
&\text{gapless}&  \:\: {\rm for}  \:\: 8.26214<\kappa<9.98933, \nonumber\\
\nu_{15/16}^c&=& 0  \:\: {\rm for}  \:\: 9.98933 <\kappa \leqslant 10.\nonumber
\label{eq:chern_15_over_16}
\end{eqnarray}
%
%


%
%
\subsection{$(p,q)=(3,2)$}
%
%

%
%
\begin{figure}[h]
\includegraphics[width=0.3\columnwidth]{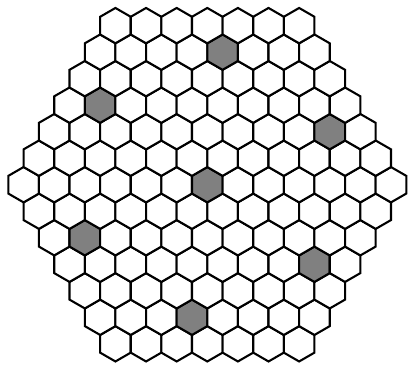} \hspace{20mm}
\includegraphics[width=0.3\columnwidth]{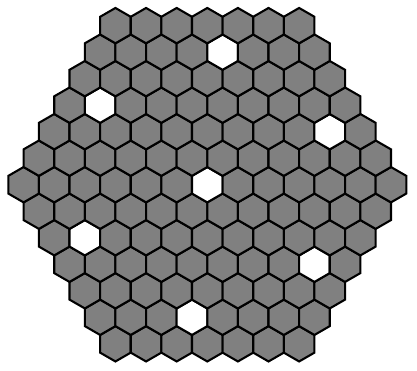}\\
\label{fig:rho_3_2}
\end{figure}
%

$\bullet$ {\em $\rho=1/19$}:
%
%
\begin{eqnarray}
\nu_{1/19}^a&=& +2 \:\: {\rm for}  \:\: 0<\kappa<0.482156, \nonumber\\
\nu_{1/19}^b&=&  -2 \:\: {\rm for}  \:\: 0.482156<\kappa \leqslant 10.\nonumber
\label{eq:chern_1_over_19}
\end{eqnarray}
%
%

$\bullet$ {\em $\rho=18/19$}:
%
%
\begin{eqnarray}
\nu_{18/19}^a&=& +1 \:\: {\rm for}  \:\: 0<\kappa<0.155448, \nonumber\\
\nu_{18/19}^b&=& +2 \:\: {\rm for}  \:\: 0.155448<\kappa<0.496882, \nonumber\\
\nu_{18/19}^c&=&  +1\:\: {\rm for}  \:\: 0.496882<\kappa<0.516295, \nonumber\\
\nu_{18/19}^d&=& -2  \:\: {\rm for}  \:\: 0.516295<\kappa \leqslant 10.\nonumber
\label{eq:chern_18_over_19}
\end{eqnarray}
%
%

%
%
\subsection{$(p,q)=(4,1)$}
%
%

%
%
\begin{figure}[h]
\includegraphics[width=0.3\columnwidth]{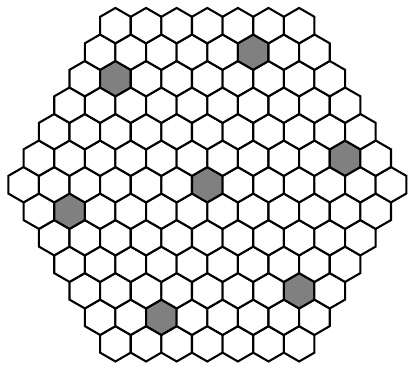} \hspace{20mm}
\includegraphics[width=0.3\columnwidth]{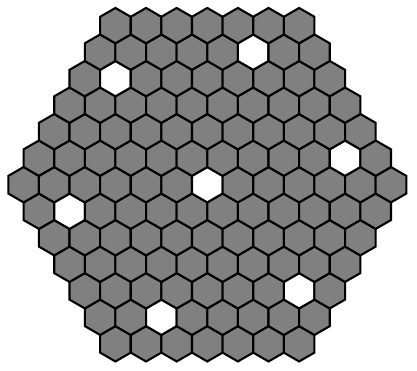}\\
\label{fig:rho_4_1}
\end{figure}
%

$\bullet$ {\em $\rho=1/21$}:
%
%
\begin{eqnarray}
\nu_{1/21}^a&=& 0 \:\: {\rm for}  \:\: 0\leqslant\kappa<0.148341, \nonumber \\
\nu_{1/21}^b&=&  +4 \:\: {\rm for}  \:\: 0.148341<\kappa<0.149193, \nonumber \\
\nu_{1/21}^c&=&  -2 \:\: {\rm for}  \:\: 0.149193<\kappa<0.150506, \nonumber\\
\nu_{1/21}^d&=&  +2 \:\: {\rm for}  \:\: 0.150506<\kappa<4.83608, \nonumber\\
\nu_{1/21}^e&=&  -2 \:\: {\rm for}  \:\: 4.83608<\kappa \leqslant 10.\nonumber
\label{eq:chern_1_over_21}
\end{eqnarray}
%
%

$\bullet$ {\em $\rho=20/21$}:
%
%
\begin{eqnarray}
\nu_{20/21}^a&=&  0\:\: {\rm for}  \:\: 0<\kappa<0.106486, \nonumber\\
\nu_{20/21}^b&=& +2 \:\: {\rm for}  \:\: 0.106486<\kappa<0.510681, \nonumber\\
\nu_{20/21}^c&=& -1 \:\: {\rm for}  \:\: 0.510681<\kappa<0.511576, \nonumber\\
\nu_{20/21}^d&=&  -2 \:\: {\rm for}  \:\: 0.511576<\kappa \leqslant 10.\nonumber
\label{eq:chern_20_over_21}
\end{eqnarray}
%
%

%
%
\subsection{$(p,q)=(5,0)$}
%
%

%
%
\begin{figure}[h]
\includegraphics[width=0.3\columnwidth]{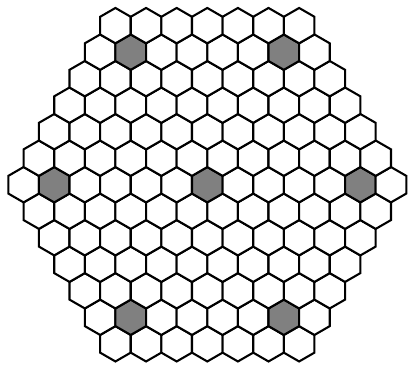} \hspace{20mm}
\includegraphics[width=0.3\columnwidth]{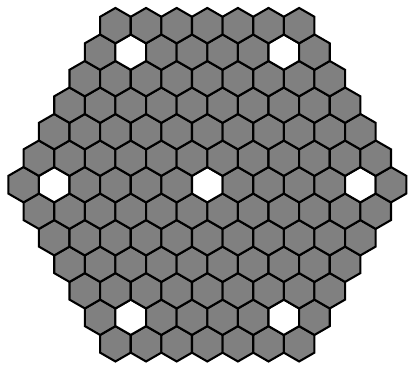}\\
\label{fig:rho_5_0}
\end{figure}
%
%

$\bullet$ {\em $\rho=1/25$}:
%
%
\begin{eqnarray}
\nu_{1/25}^a&=& +2 \:\: {\rm for}  \:\: 0<\kappa<0.056396, \nonumber\\
\nu_{1/25}^b&=&   -2\:\: {\rm for}  \:\: 0.056396<\kappa<0.079768, \nonumber\\
\nu_{1/25}^c&=&   +4 \:\: {\rm for}  \:\: 0.079768<\kappa<0.100855, \nonumber\\
\nu_{1/25}^d&=&   0 \:\: {\rm for}  \:\:   0.100855<\kappa<0.180344, \nonumber\\
\nu_{1/25}^e&=&   +4 \:\: {\rm for}  \:\: 0.180344<\kappa<0.185653, \nonumber\\
\nu_{1/25}^f&=&    -2\:\: {\rm for}  \:\: 0.185653<\kappa<0.188301, \nonumber\\
\nu_{1/25}^g&=&   +2 \:\: {\rm for}  \:\:   0.188301<\kappa<0.427832, \nonumber\\
\nu_{1/25}^h&=&   -2 \:\: {\rm for}  \:\:  0.427832<\kappa<0.456511, \nonumber\\
\nu_{1/25}^i&=&   +4 \:\: {\rm for}  \:\: 0.456511<\kappa<0.476624, \nonumber\\
\nu_{1/25}^j&=&   0\:\: {\rm for}  \:\: 0.476624<\kappa<4.58291, \nonumber\\
\nu_{1/25}^k&=&  +4 \:\: {\rm for}  \:\: 4.58291<\kappa \leqslant 10.\nonumber
\label{eq:chern_1_over_25}
\end{eqnarray}
%
%

$\bullet$ {\em $\rho=24/25$}:
%
%
\begin{eqnarray}
\nu_{24/25}^a&=&  +1\:\: {\rm for}  \:\: 0<\kappa<0.11127, \nonumber\\
\nu_{24/25}^b&=&  +2\:\: {\rm for}  \:\: 0.11127<\kappa<0.496047, \nonumber\\
\nu_{24/25}^c&=&  +1\:\: {\rm for}  \:\: 0.496047<\kappa<0.514012, \nonumber\\
\nu_{24/25}^d&=&  -2 \:\: {\rm for}  \:\: 0.514012<\kappa \leqslant 10.\nonumber
\label{eq:chern_20_over_21}
\end{eqnarray}
%
%

%
%
\subsection{$(p,q)=(3,3)$}
%
%

%
%
\begin{figure}[h]
\includegraphics[width=0.3\columnwidth]{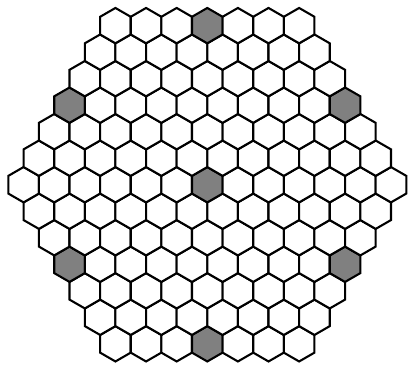} \hspace{20mm}
\includegraphics[width=0.3\columnwidth]{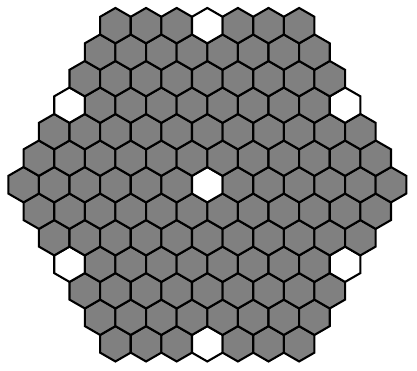}\\
\label{fig:rho_3_3}
\end{figure}
%

$\bullet$ {\em $\rho=1/27$}:
%
%
\begin{eqnarray}
\nu_{1/27}^a&=& 0 \:\: {\rm for}  \:\: 0\leqslant \kappa<0.292024, \nonumber\\
\nu_{1/27}^b&=&   +4\:\: {\rm for}  \:\:  0.292024<\kappa<0.294660, \nonumber\\
\nu_{1/27}^c&=&   -2 \:\: {\rm for}  \:\: 0.294660<\kappa<0.300077, \nonumber\\
\nu_{1/27}^d&=&   +2 \:\: {\rm for}  \:\:   0.300077<\kappa<1.04690, \nonumber\\
\nu_{1/27}^e&=&   -2 \:\: {\rm for}  \:\: 1.04690<\kappa<1.74513, \nonumber\\
\nu_{1/27}^f&=&    +4\:\: {\rm for}  \:\:     1.74513<\kappa<4.25030, \nonumber\\
\nu_{1/27}^g&=&   0 \:\: {\rm for}  \:\:   4.25030<\kappa<5.31898, \nonumber\\
\nu_{1/27}^h&=&   4 \:\: {\rm for}  \:\:    5.31898<\kappa<5.69712, \nonumber\\
\nu_{1/27}^i&=&   -8 \:\: {\rm for}  \:\:   5.69712<\kappa \leqslant 10.\nonumber
\label{eq:chern_1_over_27}
\end{eqnarray}
%
%

$\bullet$ {\em $\rho=26/27$}:
%
%
\begin{eqnarray}
\nu_{26/27}^a&=& 0 \:\: {\rm for}  \:\: 0<\kappa<0.129881, \nonumber\\
\nu_{26/27}^b&=&   +2\:\: {\rm for}  \:\:  0.129881<\kappa<0.505740, \nonumber\\
\nu_{26/27}^c&=&   -1 \:\: {\rm for}  \:\:  0.505740<\kappa<0.518925, \nonumber\\
\nu_{26/27}^d&=&   -2 \:\: {\rm for}  \:\:   0.518925<\kappa \leqslant 10.\nonumber
\label{eq:chern_26_over_27}
\end{eqnarray}
%
%

\newpage 

%
%
\subsection{$(p,q)=(4,2)$}
%
%

%
%
\begin{figure}[h]
\includegraphics[width=0.3\columnwidth]{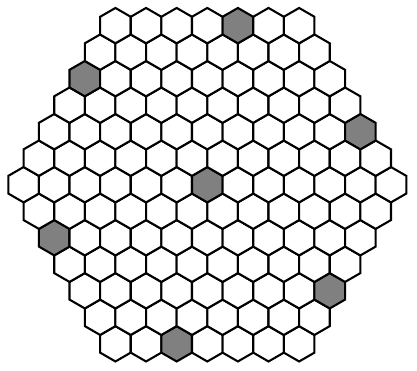} \hspace{20mm}
\includegraphics[width=0.3\columnwidth]{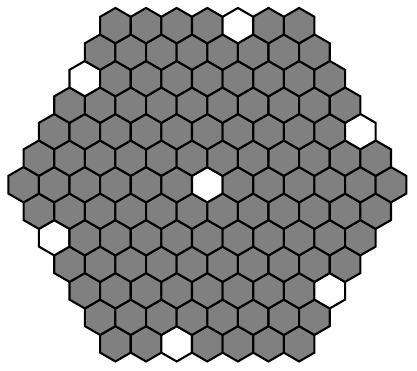}\\
\label{fig:rho_4_2}
\end{figure}
%
 
$\bullet$ {\em $\rho=1/28$}:
%
%
\begin{eqnarray}
\nu_{1/28}^a&=&  +2 \:\: {\rm for}  \:\: 0<\kappa<0.101975, \nonumber\\
\nu_{1/28}^b&=&  -2\:\: {\rm for}  \:\:  0.101975<\kappa<0.104380, \nonumber\\
\nu_{1/28}^c&=&   +4 \:\: {\rm for}  \:\:  0.104380<\kappa<0.107198, \nonumber\\
\nu_{1/28}^d&=&    0 \:\: {\rm for}  \:\: 0.107198<\kappa<3.04466, \nonumber\\
\nu_{1/28}^e&=&    +4\:\: {\rm for}  \:\:  3.04466<\kappa<3.31539, \nonumber\\
\nu_{1/28}^f&=&   +8  \:\: {\rm for}  \:\:   3.31539<\kappa<5.28623, \nonumber\\
\nu_{1/28}^g&=&   -4  \:\: {\rm for}  \:\:   5.28623<\kappa<7.24387, \nonumber\\
\nu_{1/28}^h&=&    -8\:\: {\rm for}  \:\:  7.24387<\kappa<8.67806, \nonumber\\
\nu_{1/28}^i&=&   +4 \:\: {\rm for}  \:\:    8.67806<\kappa \leqslant 10.\nonumber
\label{eq:chern_1_over_28}
\end{eqnarray}
%
%

$\bullet$ {\em $\rho=27/28$}:
%
%
\begin{eqnarray}
&\text{gapless} &  \:\: {\rm for}  \:\: 0<\kappa<0.121317, \nonumber\\
\nu_{27/28}^a&=&+2 \:\: {\rm for}  \:\: 0.121317<\kappa<0.499846 \nonumber\\
&\text{gapless}&  \:\: {\rm for}  \:\:0.499846<\kappa<0.518987, \nonumber\\
\nu_{27/28}^b&=&-2  \:\: {\rm for}  \:\:0.518986 <\kappa<9.27554, \nonumber\\
&\text{gapless}&  \:\: {\rm for}  \:\: 9.27554<\kappa \leqslant 10.\nonumber
\label{eq:chern_27_over_28}
\end{eqnarray}
%
%

\medskip 

%
%
\section{Analytical expression of the gap in the vortex-free and vortex-full sectors}
\label{app:gaps}
%
%

In this appendix, we give the exact expressions of the energy spectrum and of the gap $\Delta$ for the vortex-free and vortex-full sectors that are both defined by ${\bf A_1}={\bf n_1}$ and ${\bf A_2}={\bf n_2}$, i.e., $p=1$ and $q=0$.\\

$\bullet$ The spectrum of the Hamiltonian in the vortex-free sector ($\rho=0$) consists of two bands \cite{Kitaev06},
\be
\varepsilon({\bf k})=\pm \sqrt{|f({\bf k})|^2+g({\bf k})^2},
\ee
with 
\begin{eqnarray}
f({\bf k})&=& J (1+\rm{e}^{\rm{i {\bf k.n_1}}}+\rm{e}^{\rm{i {\bf k.n_2}}}), \nonumber \\
g({\bf k})&=&2 \kappa \big\{ \sin({\bf k.n_1})-\sin({\bf k.n_2})-\sin[{\bf k.(n_1-n_2)}]\big \}. \nonumber
\end{eqnarray}

The gap in this sector is given by
%
%
\begin{equation}
\Delta= 2  \min_{{\bf k}}\sqrt{|f({\bf k})|^2+g({\bf k})^2}=2 \min(3\sqrt{3} \: |\kappa|,|J|).
\label{eq:gap_rho_0}
\end{equation}
%
%
\medskip

$\bullet$ The spectrum of the Hamiltonian in the vortex-full sector ($\rho=1$) consists of four bands \cite{Lahtinen08,Lahtinen10},
%
%
\begin{equation}
\varepsilon({\bf k}) = \pm \sqrt{f({\bf k})\pm 2 \sqrt{g({\bf k})}},
\label{eq:gap_rho_1}
\end{equation}
%
%
with 
\begin{widetext}
%
%
\begin{eqnarray}
f({\bf k})&=&3 J^2 + 4\kappa^2 \Big[\sin^2({\bf k}.\bn_1)+\sin^2({\bf k}.\bn_2) +\cos^2({\bf k}.\bn_1-{\bf k}.\bn_2)\Big],\\
g({\bf k})&=&J^4 \Big[\cos^2({\bf k}.\bn_1)+\cos^2({\bf k}.\bn_2)+\sin^2({\bf k}.\bn_1-{\bf k}.\bn_2)\Big]+ 4\kappa^2 J^2 \Big\{4 \Big[\sin^2({\bf k}.\bn_1)+\sin^2({\bf k}.\bn_2) +\cos^2({\bf k}.\bn_1-{\bf k}.\bn_2)\Big]-3\Big\}. \nonumber \\
\label{eq:gap_rho_1}
\end{eqnarray}
%
%
\end{widetext}

Note several typographical errors in Eq.~(26) of Ref.~\cite{Lahtinen08} where this result was first derived.

The gap in this sector is given by
%
%
\begin{eqnarray}
\Delta&=&2  \min_{{\bf k}}  \sqrt{f({\bf k})- 2 \sqrt{g({\bf k})}}.
\label{eq:gap_rho_0}
\end{eqnarray}
%
%

%

\end{document}